\documentclass{elsart}

\usepackage{graphicx}
\usepackage{bm} 

\begin{document}

\begin{frontmatter}
\title{\bf Balance functions in a thermal model with resonances\thanksref{grant}}
\thanks[grant]{Research supported in part by the Polish State
Committee for Scientific Research, grant 2 P03B 059 25}  
\author[INP]{Piotr Bo\.zek},
\author[INP]{Wojciech Broniowski}, and 
\author[INP,AS]{Wojciech Florkowski}

\address[INP]{The H. Niewodnicza\'nski Institute of Nuclear Physics, \\
         Polish Academy of Sciences, PL-31342 Krak\'ow, Poland}
\address[AS]{Institute of Physics, \'Swi\c{e}tokrzyska Academy, PL-25406 Kielce, Poland}

\begin{abstract} 
The $\pi^+ \pi^-$ balance function in rapidity is computed in a
thermal model with resonances. It is found that the  correlations from
the neutral-resonance decays are important, yielding about a half of
the total contribution, which in general consist of resonance and
non-resonance parts. The model yields the
pionic balance function a few per cent wider that what follows from the recent
data for the Au+Au collisions at $\sqrt{s_{NN}}=130$~GeV.
\end{abstract}


\end{frontmatter}

\vspace{-7mm} PACS: 25.75.-q, 24.85.+p

Microscopic processes of particle production in heavy-ion collisions
conserve, in an obvious manner, the electric charge.  The {\em charge
balance function} in rapidity has been proposed as a convenient
measure of the resulting correlations between the opposite charges
\cite{baldef,Bass,jeon}. Moreover, its features can be used to
discriminate between different mechanisms of particle production. In
particular, the width of the balance function in rapidity could pin
down the time of production of the opposite-charge pair and provide
more insight on its subsequent transport in the hadronic environment
\cite{Bass}. In addition, the balance functions are closely related to 
charge fluctuations
\cite{Jeon:2000wg,Asakawa:2000wh,GM,BK,Shu,heis,Mrowczynski:2001mm}, 
whose study reveals important clues on the evolution of the system formed in the collision.
 
The balance functions analyzed by the STAR Collaboration
\cite{STARbal,phdmsu} at RHIC are defined by the formula
\begin{equation}
B(\delta,Y) = {1\over 2} 
\left\{
{\langle N_{+-}(\delta) \rangle - \langle N_{++}(\delta) \rangle \over
\langle N_+ \rangle} 
+
{\langle N_{-+}(\delta) \rangle - \langle N_{--}(\delta) \rangle \over
\langle N_- \rangle} \right\},
\label{def}
\end{equation} 
where $N_{+-}(\delta)$ counts the opposite-charge pairs which satisfy
the condition that both members of the pair fall into the rapidity
window $Y$ with the relative rapidity $|y_2-y_1|=\delta$,
whereas $N_{+}$ is the number of positive particles in the interval
$Y$. Other quantities appearing in Eq. (\ref{def}) are defined in an
analogous way. For sufficiently large rapidity interval $Y \sim Y^{\rm
max}$, the balance function of {\em all} charged hadrons is normalized to
unity \cite{Bass},
which is a condition reflecting the overall charge conservation.

The recent measurement of the balance functions by the STAR
Collaboration \cite{STARbal} showed that their widths are smaller than
expected from models discussed in Ref.~\cite{Bass} and
significantly smaller than observed in elementary particle collisions.
This issue has been recently addressed by Bia\l as \cite{Bialas}, where
the small widths of the balance functions have been explained in the
framework of the coalescence model \cite{coal}. In this paper we offer an
alternative calculation of the width of the $\pi^+ \pi^-$
balance function, based on a thermal model with resonances.  
A similar calculation has been recently performed by Pratt {\em et al.} in the 
blast-wave model \cite{Warsaw,many}.

Within our approach, the $\pi^+ \pi^-$ balance
function has two contributions, corresponding to two different
mechanisms of the creation of an opposite-charge pair. The first one
{\it (resonance contribution)} is determined by the decays of neutral
hadronic resonances. The second one {\it (non-resonance contribution)}
is related to other possible correlations among the charged
particles. We stress that in our terminology the {\em resonance}
contribution refers only to the decays of {\it neutral} resonances
which have a $\pi^+ \pi^-$ pair in the final state.  

Neutral resonances, having as their decay products a pair of the
opposite-charge pions, produce correlations between charges given
simply by the kinematics of the decay. 
In our calculation we explicitly include
\begin{equation}
K_S, \;\eta, \; \eta^\prime, \; \rho^0, \; \omega, \; \sigma, \; {\rm and} \; f_0.
\label{res}
\end{equation}
For the scalar-isoscalar channel, involving
a wide resonance, we use the formalism with the phase-shifts described in Ref.~\cite{rho}. 
The resonance contribution to the
balance function is completely defined in the framework of the thermal
model.  Neutral mesons are present in the final hadronic cocktail in
any model of the reaction. Moreover, they have been directly identified
in the invariant-mass spectra by the STAR collaboration \cite{fachini}
(see also Ref. \cite{rho}, where the STAR result is interpreted in the
framework of the thermal model used in this paper).  The parameters
which must be specified are the percentage of pions originating from
the decay of neutral resonances and the momentum distribution for such
resonances, here taken from the thermal model of Ref.~\cite{thermal}.

A different, non-resonance, mechanism for charge balancing is required
when the charged particles are produced directly in the hadronization
process, for instance during the elementary collision or the
string breakup \cite{baldef,balelem}.
 In this case the details of the process are not known,
except for the requirement of the charge conservation.  In the spirit
of the thermal model with single freeze-out \cite{thermal}, we adopt
the late-hadronization hypothesis and assume that the non-resonance
charges are balanced locally at the freezout surface, {\it i.e.}, the
two opposite-charge particles are produced at the same space-time
point with thermal velocities (defined in the local rest frame of the
fire-cylinder element). An important point is to note that a neutral
resonance ends up (with a given branching ratio) as a $\pi^+\pi^-$
pair, whereas in the non-resonance mechanism of charge balancing a
charged pion can be balanced with another charged hadron, not
necessarily a pion.

According to the above discussion, the $\pi^+ \pi^-$
balance function can be constructed as a sum of the two terms
\begin{equation}
B(\delta,Y) = B_{\rm R}(\delta,Y) + B_{\rm NR}(\delta,Y).
\label{balsum}
\end{equation}
The resonance
contribution $B_{\rm R}(\delta,Y)$ is obtained directly from the
expressions describing the phase-space of the pions emitted in a
decay. One may use here one of the results obtained in
Ref. \cite{Bialas}, {\it i.e.}, the fact that the balance functions
calculated in a neutral-cluster model do not depend on correlations
between the clusters but they are determined solely by the
single-particle distribution of the clusters and the two-particle
distribution of pions in a decay of a single cluster. Replacing the neutral
clusters of Ref.~\cite{Bialas} by the neutral resonances we immediately
obtain the two-particle rapidity distribution of the $\pi^+ \pi^-$
pairs coming from the decay of a neutral resonance
\begin{equation}
\!\!\!\!\!\!{dN^{+-}_R \over dy_1 dy_2} = \int dy d^{\,2}p_\perp \int 
d^{\,2}p_1^\perp \, d^{\,2}p_2^\perp C_\pi \, 
\left({dN_R \over dy d^2p_\perp} \right) 
\rho_{R\rightarrow \pi^+ \pi^-}\left(p,p_1,p_2\right),
\label{Npm}
\end{equation}
where $C_\pi$ indicates symbolically the presence of 
kinematic cuts for the pions.   
In a thermal approach the momentum distribution of the resonance $R$
is obtained from the Cooper-Frye formula
\begin{equation}
{dN_R \over dy d^2p_\perp} = \int d\Sigma(x) \cdot p \,
f_R\left(p\cdot u(x)\right),
\label{NR}
\end{equation}
where $f_R$ is the phase-space distribution function
of the resonance,
$u^\mu$ is the hydrodynamic flow velocity at freeze-out
defined by the constant value of the invariant time $\tau$, and
$d\Sigma_\mu$ describes a three-dimensional element of the freeze-out
hypersurface. According to Ref.~\cite{thermal}, we use the 
expansion model where $u^\mu=x^\mu/\tau$ and 
\begin{equation}
d\Sigma_\mu(x) = d\Sigma(x) \,u_\mu(x),
\label{dsigma}
\end{equation}
with $d\Sigma(x)=\tau^3 \cosh(\alpha_\perp) \sinh(\alpha_\perp)
\,d\alpha_\parallel \,d\alpha_\perp \,d\phi$. The quantities $\alpha_\parallel$
and $\alpha_\perp$ denote the longitudinal and transverse rapidity of
the fluid element, respectively, and $\phi$ is the azimuthal angle
\cite{thermal}.

The two-particle pion momentum distribution in a
two-body ($\pi^+ \pi^-$) resonance decay may be expressed by the Dirac
$\delta$ function
\begin{equation}
\rho_{R\rightarrow \pi^+ \pi^-} = {b_{\pi\pi} \over N_2}
\delta^{(4)}\left(p-p_1-p_2\right),
\label{rho2}
\end{equation}
with the normalization constant 
\begin{equation}
N_2 = \int {d^{\,3} p_1\over E_1}  {d^{\,3} p_2\over E_2}
\delta^{(4)}\left(p-p_1-p_2\right),
\label{N2}
\end{equation}
and the branching ratio $b_{\pi \pi}$.

Expressions (\ref{Npm})-(\ref{N2}) may be easily generalized to the case of
three-body decays where the charged $\pi^+ \pi^-$ pair is produced
together with an extra neutral pion. With the typical
technical assumption of a constant transition matrix element we have
\begin{equation}
\rho_{R\rightarrow \pi^+ \pi^- \pi^0} = {b_{\pi\pi\pi} \over N_3}
\int {d^{\,3} p_3\over E_3} \,
\delta^{(4)}\left(p-p_1-p_2-p_3\right),
\label{rho3}
\end{equation}
where the normalization constant is given by the formula
\begin{equation}
N_3 = \int {d^{\,3} p_1\over E_1}  {d^{\,3} p_2\over E_2}
{d^{\,3} p_3\over E_3}
\delta^{(4)}\left(p-p_1-p_2-p_3\right). 
\label{N3}
\end{equation}

The non-resonance correlations arise from 
pions emitted directly from the firecylinder or those produced in the
decays of resonances other than (\ref{res}).  In contrast to the correlations between
the resonance pions (fully determined by the model), the correlations
between the non-resonance pions are not specified by the model and we
have to propose their form.  We assume that the creation of an
opposite-charge pair occurs locally in the fire-cylinder. It means that
the two charges have the same longitudinal and transverse collective
velocity, however their relative momentum is determined by the local
thermal momenta of both particles \cite{Bass,scott2}.  This process is
described by the following two-particle rapidity distribution for
correlated pairs
\begin{eqnarray}
\label{thermalcorr}
{dN^{+-}_{NR} \over dy_1 dy_2} & = &  A \int d^{\,2}p_1^\perp 
d^{\,2}p_2^\perp C_\pi \, 
\nonumber \\
& & \times \int d\Sigma(x) 
p_1 \cdot u(x) f^\pi_{NR}\left(p_1\cdot u(x)\right)
p_2 \cdot u(x) f^\pi_{NR}\left(p_2\cdot u(x)\right).
\label{NpmNR}
\end{eqnarray}
Here $f^\pi_{NR}$ is the phase-space distribution function of the
non-resonance pions and the normalization constant $A$ in
Eq.~(\ref{thermalcorr}) is obtained from the condition
\begin{equation}
\int  dy_2  \left({dN^{+-}_{NR} \over dy_1 dy_2} \right)
= {dN^\pi_{NR} \over dy_1}.
\end{equation}
We note that a charge can be produced in the fire-cylinder directly as
a pion, a resonance decaying eventually into a charged pion, or as a
different charged hadron. The last case is not registered in the
$\pi^+\pi^-$ balance function, and leads to a reduction of its norm.
The first two cases can be effectively taken into account by combining
in the distribution $f^\pi_{NR}$ the direct pions  and the
charged pions originating from the decays of the resonances other than
(\ref{res}). This has a
softening effect on the pion spectra and leads to a slight reduction of the
width of the balance function as compared to the case with direct
pions only.  For both the resonance and the non-resonance production
of correlated opposite-charge pairs we assume that after the
freeze-out the momentum distribution of pions is not modified
by rescattering. This picture is consistent with the single-freezout
thermal model we use.

Knowing the two-particle distribution functions (\ref{Npm}) and
(\ref{NpmNR}), we calculate the balance function of the resonance
pions from the expression
\begin{equation}
B_{\rm R}(\delta) 
= {1 \over
N_\pi } \sum_R \int dy_1 dy_2 C_\pi \left({dN^{+-}_R \over dy_1 dy_2} \right)
\delta(|y_2-y_1|-\delta),
\label{balres}
\end{equation}
where the sum over $R$ includes the resonances (\ref{res}).
Similarly, for the non-resonance
contribution (marked by the tilde here, which indicates the neglect of
the balancing from other hadrons) we obtain
\begin{equation}
{\tilde B}_{\rm NR}(\delta) 
= {1 \over
N_\pi } \int dy_1 dy_2 C_\pi \left({dN^{+-}_{NR} \over dy_1 dy_2} \right)
\delta(|y_2-y_1|-\delta).
\label{balres2}
\end{equation}
In the above formulas $N_\pi=(N_{\pi^+}+N_{\pi^-})/2$ is one half of the total number of 
charged pions observed in the acceptance window of rapidity,
\begin{equation}
N_{\pi}=\int dy_1 d^{\,2}p_1^\perp \left({dN^\pi \over dy_1 d^2p_1^\perp}
\right) C_\pi. 
\label{Npi}
\end{equation}
We note that $N_\pi$ is the sum of the number of non-resonance pions
$N^\pi_{NR}$ and the number of  resonance pions $N_{\rm R}^\pi$. As a
consequence, {\em in the absence of cuts}, 
the two contributions to the pion balance functions are
normalized to the relative weight of the corresponding pion-pair
production processes. More explicitly,
\mbox{$\int_0^{Y_{\rm max}} d\delta B_R(\delta)=N_{\rm R}^\pi/N_\pi$}, and 
\mbox{$\int_0^{Y_{\rm max}} d\delta \tilde B_{NR}(\delta)=N^\pi_{NR}/N_\pi$}.
Taking into account the fact that some of the
thermal pions are balanced by other charged hadrons, the final
expression for the pion balance function is
\begin{equation}
B(\delta)=B_{\rm R}(\delta)+\frac{N^\pi_{\rm NR}}{N_{\rm charged}
-N_{\rm R}^\pi}
{\tilde B}_{\rm NR}(\delta),
\end{equation}
where $N_{\rm charged}=(N_++N_-)/2$ is a half of the total
multiplicity of the charged (positive plus negative) particles. The
factor $N^\pi_{\rm NR}/(N_{\rm charged}-N_{\rm R}^\pi)$ is the ratio of
the non-resonance pions to all charged particles, with the exclusion
of the pions coming from resonances, which have already been accounted
for in $B_{\rm R}(\delta)$.  With the parameters of the single
freezout thermal model \cite{thermal} one finds $N^\pi_{NR}/(N_{\rm
charged}-N_{\rm R}^\pi)=0.68$.

\begin{figure}[tb]
\begin{center}
\includegraphics[width=12cm]{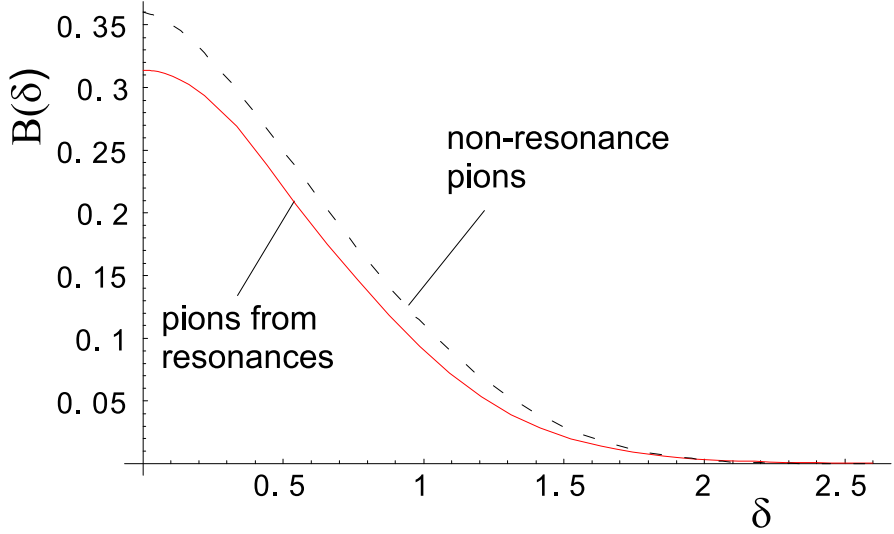}
\end{center}
\caption{Contributions to the balance function from the neutral
resonances (solid line) and from the non-resonance pions (dashed
line), plotted as a function of the rapidity difference of the two
pions. The geometric parameters of the model satisfy $\rho_{\rm max}/\tau=0.9$, which 
corresponds to $\langle \beta_\perp \rangle=0.5$.}
\label{brnr}
\end{figure}

Our model results for the two (resonance and non-resonance)
contributions to the balance function are shown in
Fig. \ref{brnr}. One can observe that the balance function for pions
originating from the neutral-resonance decays is very similar to that
obtained for the non-resonance contribution. In general, the thermal
pion balance function \cite{Bass,jeon,scott2} is known to be wider
than the experimental balance function measured by the STAR
collaboration \cite{STARbal}. The narrowing of the balance function
may be achieved by imposing the strong transverse flow, however, this
effect is not sufficient to explain the data, unless unreasonably low
values of the temperature are chosen.  The non-resonance pions in our
calculation give a relatively wide balance function, since we use
the high freeze-out temperature of $165$~MeV and an average
transverse velocity $\langle \beta_\perp \rangle=0.5$.  
The pion pairs originating from the decays of neutral resonances
lead to a similar balance function. As a result, the sum
of the two is a few per cent wider than the experiment.  The norms, including
the kinematic cuts of $|\eta_\pi|<1.3$ and $p_\perp^{(1,2)}>0.1$~GeV,
are \mbox{$\int d\delta B_R(\delta)=0.25$}, and \mbox{$\int d\delta
B_{NR}(\delta)=0.29$}, with the total reaching 0.53. The number is
less than one, which reflects the presence of cuts, and the mentioned
effect of balancing of non-resonance pions by other charged particles.

\begin{figure}[tb]
\begin{center}
\includegraphics[width=14.2cm]{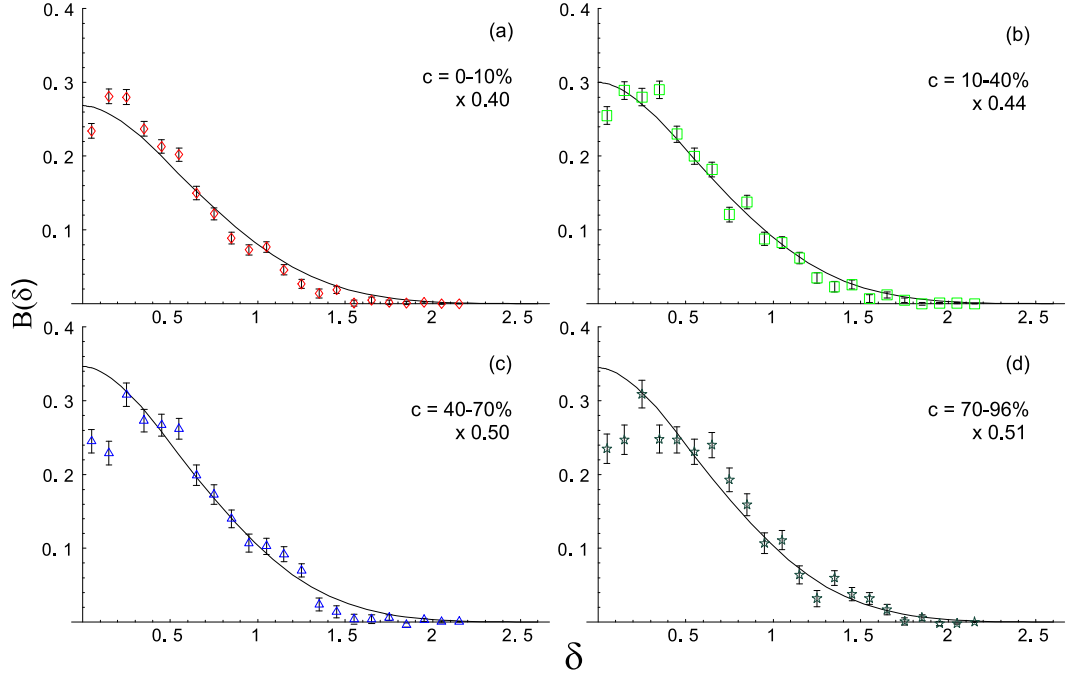}
\end{center}
\caption{Balance functions for the pions in the thermal model
calculated for four different centrality classes and compared to the
experimental data of Ref.~\cite{STARbal}. The normalization, affected
by the detection efficiency, is adjusted a posteriori, as explained in
the text. The resulting normalization factor, obtained by a $\chi^2$
fit in the range $0.2 < \delta <2.2$, is listed near the plot
labels.}
\label{thdat}
\end{figure}

In Fig. \ref{thdat} we show our full result for the balance function
for four different centrality windows compared to the STAR experimental
data. The geometric (expansion) parameters of the thermal model are
taken in such a way that their ratio $\rho_{\rm max}/\tau$ is fixed
and equal to 0.9, for each centrality. The dependence on the $\tau$
parameter vanishes, since it appears as a factor $\tau^3$ in both the
numerator and denominator of Eq.~(\ref{def}). In each case the
normalization of the balance function has been adjusted arbitrarily,
due to our lack of knowledge of the experimental efficiency \cite{phdmsu}. The
resulting factors are given below the plot labels, and vary from 0.40
to 0.51. The basic conclusion is that the model and the
experiment agree quite reaonably for the shape of $B(\delta)$. 
In fact, the model agrees best for the most periferic data, where the 
measured balance function is wider. We note that the
dips of the balance functions observed at small values of $\delta$ are
caused by the Bose-Einstein correlations and cannot be explained by
our model, unless such correlations are incorporated as an additional
effect.

The mean values of $\delta$ are presented
in Table 1 independently for the resonance and non-resonance
contributions, and also for the total. 
For comparison with the experimental values \cite{STARbal}, we present 
the results for the the mean  width $\langle \delta \rangle$ with
the exclusion of the region $\delta< 0.2$. 
In addition, in Table 1 we show the results
obtained for $\rho_{\rm max}/\tau=0.8$, which corresponds to a smaller
value of the transverse flow. 

\begin{table}[b]
\begin{center}
\begin{tabular}{|c|c||c|c|c|}
\hline
$\rho_{\rm max}/\tau$ & $\langle \beta_\perp \rangle$ & $ \langle \delta \rangle_{\rm res}$ & 
 $ \langle \delta \rangle_{\rm therm}$ & $ \langle \delta \rangle_{\rm tot}$ \\ \hline
0.8 & 0.46 & 0.65 (0.52) & 0.68 (0.54) & 0.66 (0.53) \\ 
0.9 & 0.50 & 0.65 (0.51) & 0.67 (0.53) & 0.66 (0.52) \\
\hline
\end{tabular}
\end{center}
\caption{The  widths of the pionic balance functions in rapidity for
different geometry/flow parameters. The numbers for $\langle \delta
\rangle$ are obtained with the range $0.2 < \delta < 2.6$, while the
values in parenthesis are for the range $0 < \delta < 2.6$.}
\end{table}

The obtained model width of the pion balance function is consistent with
the STAR collaboration results for the $70-96\%$ centrality bin, where
$\langle \delta \rangle = 0.664\pm.029$. For the central events ($c=0-10\%$) 
the experiment gives $\langle \delta \rangle = 0.594\pm 0.019$,  
for the mid-central ($c=10-40\%$) $\langle \delta \rangle = 0.622\pm 0.020$, 
and for the mid-periferal ($c=40-70\%$) 
$\langle \delta \rangle = 0.633\pm 0.024$, which is few per cent larger than the model result. 
The dependence of the
width of the balance function on centrality cannot be reproduced 
in our approach by varying the transverse flow within limits consistent with the
single-particle spectra.

We summarize our main points:

\begin{enumerate}

\item An important source of the correlation between the
opposite-charge pions is the decay of neutral resonances.  
The decay of the neutral resonances gives about half of
the pion pairs in the rapidity window considered.

\item The rest of the pion pairs has been assumed to be emitted
locally from the freeze-out surface \cite{Bass,jeon,Bialas,scott2}.
The balance function for such pairs is relatively wide 
since the temperature at the freeze-out surface is large
and the average transverse flow is small. Hence, this mechanism of local
charge balancing in the fireball cannot explain the data by itself.

\item By summing the contribution of the two mechanisms of charge
balancing we obtain a pion balance function with the shape similar to
the periferic data of Ref.~\cite{STARbal}.

\item The normalization of the balance function obtained from the
model is significantly larger than the experimental value (a factor of
2 - 2.5) because of the effect of a limited detector efficiency 
and acceptance that
we are not able to take into account. We include only the sharp
kinematic cuts in pseudorapidity and transverse momentum to simulate
the acceptance window of the STAR experiment.

\item The limited detector efficiency and acceptance may also affect the 
width of the balance function. 

\end{enumerate}

\section*{Acknowledgements}
We are grateful to Andrzej Bia\l{}as and Scott Pratt for
useful discussions.


\begin{thebibliography}{99}

\bibitem{baldef} D.~Drijard et al., Nucl. Phys. {\bf B 155} (1979) 269;
Nucl. Phys. {\bf B 166} (1980).

\bibitem{Bass} S.~A.~Bass, P.~Danielewicz, and S.~Pratt,
  Phys. Rev. Lett. {\bf 85} (2000) 2689.

\bibitem{jeon} S.~Jeon and S.~Pratt, Phys. Rev. {\bf C 65} (2002) 044902.

\bibitem{Jeon:2000wg}
S.~Jeon and V.~Koch,
Phys.\ Rev.\ Lett.\  {\bf 85} (2000) 2076.

\bibitem{Asakawa:2000wh}
M.~Asakawa, U.~W.~Heinz, and B.~M\"uller,
Phys.\ Rev.\ Lett.\  {\bf 85} (2000) 2072.

\bibitem{GM}
M.~Ga\'zdzicki and S.~Mr\'owczy\'nski,
Z. Phys. {\bf C 54} (1992) 127.

\bibitem{BK} A. Bialas and V.  Koch, Phys. Lett. {\bf B 456} (1999) 1. 

\bibitem{Shu}
E.~V.~Shuryak,
Phys. Lett. {\bf B 423} (1998) 9.

\bibitem{heis} H. Heiselberg and A. D. Jackson, Phys. Rev. {\bf C 63}
  (2001) 064904.

\bibitem{Mrowczynski:2001mm}
S.~Mr\'owczy\'nski,
Phys.\ Rev.\  {\bf C 66} (2002) 024904.

\bibitem{STARbal} J. Adams et al., STAR Collaboration,
  Phys. Rev. Lett. {\bf 90} (2003) 172301; 
http://www.star.bnl.gov/STAR/sds\_l/all\_l/physicsdatabase/17/data.html .

\bibitem{phdmsu} M. B. Tonjes, PhD thesis, Michigan State University (2002).

\bibitem{Bialas} A.~Bialas, hep-ph/0308245.

\bibitem{coal}
T.~S.~Biro, P.~Levai, and J.~Zimanyi,
Phys.\ Lett. {\bf B 347} (1995) 6.

\bibitem{thermal} W.~Broniowski, and W.~Florkowski,
  Phys. Rev. Lett. {\bf 87} (2001) 272302; 
Phys. Rev. {\bf C 65} (2002) 024905; 
W.~Broniowski, A.~Baran, and W.~Florkowski, Acta Phys. Pol. {\bf B 33} (2002) 4235.

\bibitem{Warsaw} S. Pratt, talk presented at Second Warsaw Meeting 
on Particle Correlations and Resonances 
in Heavy Ion Collisions, Warsaw, 15-18 October 2003
(http://hirg.if.pw.edu.pl/meeting/oct2003/talks/pratt/ScottWarsaw.ppt).


\bibitem{many} S. Cheng, C. Gale, S. Jeon, S. Petriconi, S. Pratt, M. Skoby, 
V. Topor Pop, Q.~H.~Zhang, nucl-th/0401008.

\bibitem{rho} W.~Broniowski, W.~Florkowski, and  B.~Hiller,
  Phys. Rev. {\bf C 68} (2003) 034911. 

\bibitem{fachini} J. Adams et al., STAR Collaboration,
  nucl-ex/0307023; P.~Fachini, STAR Collaboration,
nucl-ex/0305034.

\bibitem{balelem}M.~Althoff {\it et al.},  TASSO Collaboration,
Z.\ Phys.\  {\bf C 17} (1983) 5;
I.~V.~Azhinenko {\it et al.}, 
 EHS-NA22  Collaboration,
Z.\ Phys.\  {\bf C 43} (1989) 37;
H.~Aihara {\it et al.},  TPC/Two Gamma Collaboration,
Phys.\ Rev.\ Lett.\  {\bf 53} (1984) 2199.

\bibitem{scott2}
S.~Pratt and S.~Cheng,
Phys. Rev.  {\bf C 68} (2003) 014907

\end{thebibliography}
\end{document}